# SwitchAgg: A Further Step Towards In-Network Computation

Fan Yang, Zhan Wang, Xiaoxiao Ma, Guojun Yuan, Xuejun An
Institute of Computing Technology
Chinese Academy of Sciences
Beijing, China
{yangfan, wangzhan, maxiaoxiao, yuanguojun, axj}@ncic.ac.cn

## ABSTRACT

Many distributed applications adopt a partition/aggregation pattern to achieve high performance and scalability. The aggregation process, which usually takes a large portion of the overall execution time, incurs large amount of network traffic and bottlenecks the system performance. To reduce network traffic, some researches take advantage of network devices to commit in-network aggregation. However, these approaches use either special topology or middle-boxes, which cannot be easily deployed in current datacenters.

The emerging programmable RMT switch brings us new opportunities to implement in-network computation task. However, we argue that the architecture of RMT switch is not suitable for in-network aggregation since it is designed primarily for implementing traditional network functions.

In this paper, we first give a detailed analysis of in-network aggregation, and point out the key factor that affects the data reduction ratio. We then propose SwitchAgg, which is an in-network aggregation system that is compatible with current datacenter infrastructures. We also evaluate the performance improvement we have gained from SwitchAgg. Our results show that, SwitchAgg can process data aggregation tasks at line rate and gives a high data reduction rate, which helps us to cut down network traffic and alleviate pressure on server CPU. In the system performance test, the job-completion-time can be reduced as much as 50%.

## 1 Introduction

A large amount of distributed applications adopt a partition/aggregation pattern [1] to achieve high performance and scalability. This pattern covers a wide range of data-intensive frameworks, including big data analytics [2], graph processing [3], machine learning [4], and real time stream processing [5, 6, 7, 8]. Generally, these frameworks distribute a large input data set over many worker servers and each server computes on its data independently. Then the partial results of each worker will be aggregated to generate the final result.

The aggregation process, which usually takes a large portion of the overall execution time, brings us a great challenge. When workers send their intermediate results for further aggregation, a large volume of data (typically tens of gigabits [4]) is injected into the network in many-to-few pattern, which can easily saturate the in-bound link of the receiver and further decrease the performance. For example, in Facebook map/reduce jobs [9], network transfer is responsible for 33% of the execution time of the jobs, and in 16% of these jobs, network transfer occupies more than 70% of the execution time. Another work [10] points out that, network transfer is and will continue to be the bottleneck of distributed neuro network training. Thus cutting down the network traffic is the key factor to improve the overall performance.

In order to deal with this problem, various countermeasures have been proposed. One of these approaches, known as in-network aggregation [11], is the most appealing solution which assigns the aggregation tasks to the network itself. This idea is attractive for two reasons. First of all, traffic is reduced when traversing the network devices, which not only alleviates the in-bound problem but also helps to release network congestion. Second, network devices commit on-path data reduction, which relieves stress on the host CPU. Unlike some previous works [12] which offload reduction tasks to NIC, this method goes a step further, since the reduction operation is distributed among different network devices, we can easily achieve high parallelism of data reduction. Preliminary works [11, 13, 14] have shown that this approach can obtain a high data reduction rate. However, since the network device (typically switch or router) usually has a fixed function set, this approach is implemented using either special topology [13] or custom middle-box [11], which needs to replace the infrastructure of the datacenter. Another choice [15] is to use dedicated hardware, such as Mellanox SwitchIB-2, but it only focuses on scientific applications, typically MPI primitives.

The recently proposed RMT switch architecture [16, 17, 18] brings us new opportunities for in-switch aggregation. The RMT architecture has a multi-stage pipeline where packets flow at line rate. Each stage is composed of a match unit and an action unit, coupled with a memory cluster (SRAM and TCAM), allowing for matching and modifying the header fields. Recent work [14] has implemented aggregation function in RMT switch. It takes aggregated data as several Key-Value pairs, and aggregates the value of the same key when data flows through the switch. The preliminary result has proved that, with little work of development, we can significantly reduce the traffic and achieve an increase in overall performance.

However, this solution relies on several assumptions and has some drawbacks. First of all, it assumes that the format of the data is known beforehand, i.e. launching a new aggregation task which has a different data format needs to recompile all the switches in the network. Besides, all the data should be of fixed length, for example, it requires all the Key-Value pairs to be of fixed 24B which facilitates header parsing. Those Key-Value pairs whose lengths are smaller will be padded with zero which incurs extra traffic. In addition, data should be encapsulated into a fixed-length packet header, which lacks flexibility. Besides, it has a limitation of key variety to be about 16K, which fits properly into the switch memory, but far less than the real use case. Last but not least, current P4 switches are expected to handle packet has a length of only around 200B ~ 300B [14], which generates more packets and incurs extra header overhead.

Based on the observations above, we argue that the existing RMT architecture is not fit for in-network aggregation for the following two reasons:

*1).Limited memory capacity.*

The typical size of on-chip memory of the commercial programmable switching chip[16] is not enough to accommodate all the different keys we need to aggregate (which usually reaches several gigabits). Even we can take an aggressive approach to forward the data which exceeds the capacity limitation to the next hop, our later experiment will show that, with the increasing variety of different keys, the memory capacity is the dominant factor that limits the data reduction rate.

*2). Inflexible header parsing.*

Once the switch has been configured, the header parser is set to a finite state machine, which is primarily designed for adding new protocols to the switch. This mechanism of header parsing is useful and adequate for several network applications such as in-network cache [19, 20], consensus protocol [21] and network monitoring [22]. However, it is not suitable for the aggregation task. When the packet carries various-length key-value pairs, the parser can hardly deal with it. Even if we constrain the data to a fixed length in one task, when launching another job, we will rewrite the program and reconfigure all the switches, which needs a great effort.

These limitations pose several challenges for in-switch aggregation:

1).The switch should be able to process aggregated packets at line rate.

2).Multiple applications with different requirements can share the same deployment without too much modification.

3).It should maintain a high data reduction ratio irrespective of data amount.

To meet these challenges, we propose SwitchAgg, a switch architecture which is well-suited for in-network aggregation. SwitchAgg consists of three parts: 1). A payload analyzer which can handle different-length key-value pairs; 2). Multiple processing elements which execute both normal forwarding task and aggregation job at line rate; 3). A two-level memory hierarchy which is composed of a private SRAM and a shared back-end DRAM, aiming at overlapping computation and memory accessing. Unlike NPU-based solution which usually suffers a lot from cache misses [20], our carefully designed memory hierarchy and overlapping mechanism ensures that there is no penalty when cache miss happens, which ensures a high throughput.

Our prototype is implemented on a NetFPGA platform. In order to measure the overall improvement in application level, we also develop a MapReduce-like framework and a light middle-layer which facilitates the application to transfer data. The experimental results have shown that, our prototype can handle aggregation task at line rate and offers a high data reduction rate even when there is a large variety of data. At the application level, we show an performance improvement of 50%.

We make four specific contributions:

1). Give a detailed analysis of the limitation and drawbacks of current programmable switch when dealing with in-network aggregation;

2). Implement a prototype of SwitchAgg to give a concrete example of our design;

3). Use our prototype to quantify the benefits of in-network aggregation in a MapReduce-like system;

4). Present a number of building blocks for future switch design and take a step further towards a wider range of in-network computing.

## 2 Background and Motivation

In this section, we first give a brief review of the idea of in-network aggregation with definitions of some related terminology, and justify the feasibility of this idea (§2.1). Then we give concrete examples and detailed analysis to explain why existing programmable switch fall short in dealing with the aggregation problem and motivate our design (§2.2).

### 2.1 In-network Aggregation

An in-network aggregation system can be roughly divided into two parts. ***Aggregation nodes*** and ***aggregation tree***. An aggregation node is a logical concept, which can be implemented on a network device(e.g. a middlebox[11], a software router[14] or a switch[15]), it aggregates data carried by the downstream flows and forwards the intermediate results to the upper-level node. The proportion that the output data occupies in the input data is called reduction ratio, this concept can be applied to a single aggregation node or the whole aggregation tree. All the aggregation nodes work coordinately and they form an aggregated tree which is pre-configured by the developer. Figure 1 gives an example of a MapReduce task, which contains seven Mappers and one Reducer, and illustrates the logical view of aggregations tree of these participants.

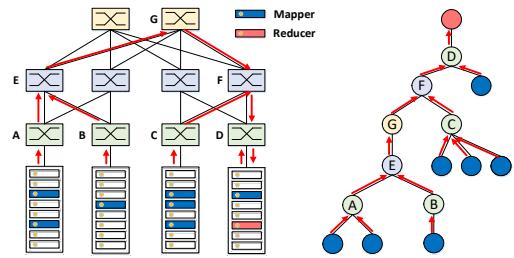

Figure 1. The aggregation tree and aggregation node

The aggregation functions have several features that are well-suited for in-network implementation. First, these aggregation operations typically consist of simple arithmetic or logical operations, which facilitates parallel execution. Second, which is the most important, these aggregation operations usually meet the demands of commutability and associativity (such as SUM, MAX, TOP-k), which implies that they can be executed separately on different parts of the data, with no regard to the execution order.

## 2.2 Programmable Switch for Aggregation

The prevailing architecture for programmable switches is the Reconfigurable Match-Action Table (RMT) [16]. RMT uses a pipeline of match-action stages, which can be programmable and reconfigured. Developers use P4 language to specify the format of headers and the actions they need to perform. These features make programmable switch a potential choice for in-network aggregation since it can be integrated into the existing data center infrastructure seamlessly.

DAIET[14] is a representative in-network aggregation system which is implemented on a software RMT model [16]. It defines a new type of packet header which is consisted of several Key-Value pairs of identical length. When the switch receives a packet, it extracts the Key-Value pairs from the header and matches the Key against the lookup table which is placed in SRAM or TCAM. If the table entry is empty, it stores the Key with the Value. Otherwise, if the Key is found in the table, the action unit aggregates the value of this key with the previously stored value, or forward this key to next-hop if not found.

Owing to some architectural limitations, such as table size, DAIET makes several important assumptions to facilitate its implementation. We point out that these assumptions are far from the real cases which are not well-suited for implementation on RMT switch. We detail these limitations in two parts.

### 2.2.1 Inflexible Header Parser

Through quantitative analysis, we observe two major problems of the header parser.

*Extra traffic:* Due to the characteristics of the header parser, we notice two factors will contribute to extra network traffic.

1). Since all the key-value pairs are encapsulated into the header parser, they should be of an identical structure. For example, DAIET designated that all the key-value pairs should be fit into <16B-Key, 4B-Value> format. Those key-value pairs whose lengths are less than 20B will be padded with zero. Consider the following case that a packet has a length of 200B which contains 10 KV-pairs. Suppose the average length of these KV-pairs is 10B, we need to inject about 50% more traffic into the network.

We model the extra amount of data that needs to be transferred as follows: we assume that a RMT switch can handle packets that has a maximum length of M Bytes, each packet contains fixed- and identical-length key-value pairs of N Bytes, and the actual length of each key-value pair is $P_i$ (1 <= i <= $\lfloor M/N \rfloor$ and 1 <= N <= M). The extra traffic ratio we need to transfer for a simple packet is given by the following formula.

$$T = M / \left( \sum_{i=1}^{n} P_i \right) \quad (1)$$

In an extreme case, where M is 200, N is 20 (i.e. the longest Key-Value pair is 20B), and $P_i$ = 1, we need to transfer nearly 7 times more data for a single packet.

2). The extra traffic is correlated with another factor. Typically, the RMT switch can handle a packet of 200B at most [14], when transferring a certain amount of data, it may generate more packets which incurs a relatively expensive header overhead. For a certain amount of data D, the total bytes we need to inject into the network is:

$$T = D + \lfloor D/M \rfloor * H \quad (2)$$

Where M is the maximum data a packet can carry with and H is the length of the protocol header overhead (58B for a TCP/IP packet). Specially, RMT constrains the packet length to be within 200B, while a traditional TCP packet can accommodate ~1500B payload. Hence we can figure out the extra header overhead ratio is 25.3%, which may even offset the benefits we get from in-network aggregation.

*Inflexibility:* The primary goal of RMT architecture is to deploy new protocols without modifications to existing hardware. When deploying a new protocol, all we need to do is to write a new P4 program and recompile the switch. It is well-suited for new protocols verification whose changing rate is several days or even months. However, in-network aggregation has a totally different requirement. First of all, the workloads are rapidly changing, in the previous example, we constrain the Key-Value pair to be of 20B, however, when the task requires larger key-value pairs, we need to recompile all the switches. Besides, once the switch is configured, it can only provide service for one job, another job with different data format cannot take advantage of this configuration.

### 2.2.2 Memory Limitation

Another problem is that the memory capacity constrains the data reduction ratio. DAIET assumes that the variety of different keys is no more than 16K, which is an ideal size to be fit into the switch memory. However, in reality, the number of Key-Value pairs can be much larger. For example, some graph processing systems [23] use Key-Value pairs to represent the properties of vertexes and edges, whose number can be as many as several millions. In this case, most of the data cannot be aggregated in one hop and will be forwarded to next hop. We argue that this limitation of memory capacity severely constrains the data reduction ratio. We first model this problem and then verify our theory with experiments. Before modeling, we propose a theorem to simplify our further experiments and design.

**THEOREM 2.1** The reduction ratio of an aggregation node which receives multiple flows is the same as merging these flows into one and transferring it through.

Based on this theorem, we model this problem as follows, one flow traverses through a programmable switch and the switch aggregates the data that flow carries. Suppose the Key-Value pair has an average length of L, the data amount is M, the memory capacity is C (M and C are measured in the units of L), the Key variety is N (M >= N), and the data is evenly distributed among N variety. The reduction rate can be calculated as follows:

$$R = \begin{cases} 1 - \dfrac{N}{M} & if\ (N \le C) \\ \left(\dfrac{1}{N} - \dfrac{1}{M}\right) \cdot C & if\ (N > C) \end{cases} \quad (3)$$

The highest reduction ratio is bounded to C / N. To verify out theory, we conduct an experiment. A hardware-based packet generator continuously produces packets which contains different key-value pairs with identical length of 20B. And a processing engine is responsible for aggregating data that it provided by the packet generator. The memory capacity is constrained to 16MB. And the total data amount is 1GB, which represents a median data amount that a worker node needs to transfer in a MapReduce system [9]. Figure. 2(a) shows the impact of Key variety on reduction ratio.

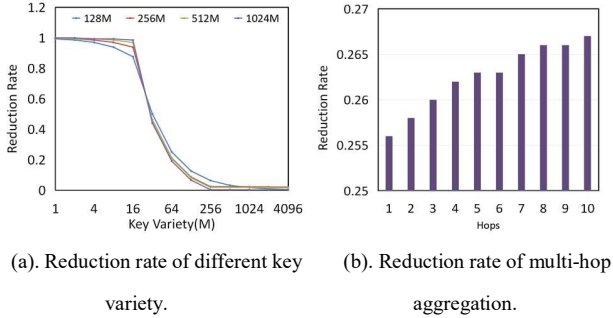

(a). Reduction rate of different key variety.

(b). Reduction rate of multi-hop aggregation.

Figure 2. Reduction rate in different cases

From the experiments, we have two important observations. 1). When the memory is large enough to accommodate all the varieties of different keys, we can easily obtain a reduction rate higher than 80%, however, 2). when the key variety increases and exceed the memory capacity, we suffer a cascading of the reduction rate. When the key variety is one order of magnitude of the memory capacity, the reduction rate is below 10%, and even below 1% when we have 4G different Keys.

Another question is, will it help when aggregation happens in a multi-level aggregation tree? The answer depends on the characteristics of the data. i.e. whether the data is evenly distributed among different key varieties or not. This problem can be described in the following theorem.

**THEOREM 2.2** When data is evenly distributed among different key varieties, the results of multi-hop aggregation is exactly the same to single-hop aggregation; When data is non-uniformly distributed, the reduction ratio of multi-hop aggregation has the same upper- and lower-bound of the singe-hop aggregation.

We conduct another experiment which connects more devices in a streamline. If the upper-level switch cannot accommodate all the keys, it will forward these keys to the downstream switch for further aggregation. We choose evenly-distributed data set and the key variety is 64M, the total data amount is 1GB with a memory capacity of 128MB. We expect a higher reduction rate when we use several switches. However, the outcome is not as expected. As shown in Figure 2(b), we see that the increasing hops of processing stages does not help a lot, hence we conclude that the single-hop memory capacity is the key factor that limits the data reduction rate.

The above analysis and experiments motivate us to design a new switch architecture which can better deal with the in-network aggregation task. We propose several requirements for our design: 1). The architecture should be able to handle various types of data format in order to support several different aggregation tasks; 2). It must be accompanied with a relatively larger memory capacity to achieve better reduction ratio, without incurring latency or degrading throughput. The following sections give a detailed description of our design.

## 3. SwitchAgg Overview

SwitchAgg is a new switch architecture for implementing in-network data aggregation. It processes packets at line rates and provides a high data reduction ratio, which helps to reduce the network traffic and alleviate computational pressure on end-host CPU. Figure 3 shows the overall architecture of SwitchAgg, which consists of a switch, a controller and end-hosts.

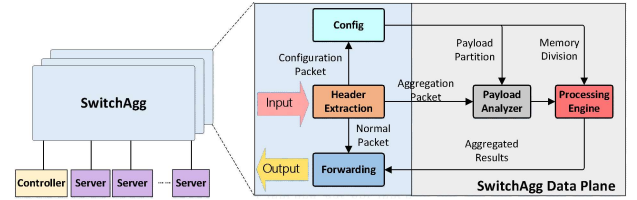

Figure 3. SwitchAgg Overview

**Switch.** The switch is the main component of SwitchAgg. It is responsible for 1). Aggregating flows coming from different ports, and 2). Flushing aggregation results to the next hop. 3). Forwarding normal packets.

The header extraction module firstly examines the packet header to determine whether a packet is needed to be aggregated. If the packet is marked as an aggregation packet, it will be forwarded to the payload analyzer for further processing. Otherwise, the packets will be processed by the forwarding module based on its L2/L3 information in the traditional way.

The configuration module accepts the control information from the controller, and configures the switch for new aggregation tasks, which contains information of the logical topology of an aggregation tree, the number of flows needs to be aggregated, etc. The payload analyzer with the processing engine is the core part to implement the aggregation function. The former one accepts payload which is formatted as Key-Value pairs and transfers these pairs to different processing engines based on their lengths. The processing engines are responsible for aggregating values of the same key. They first store a pair into the memory, when another pair came, it looks up the table, if the key is found, it aggregates the value, otherwise it replaces the key. Each processing engine is dedicated for processing fixed-length Key-Value pair, which facilitates the key lookup and replacement, we will detail this design later.

*Controller.* The controller is primarily responsible for configuring the control plane of the switch. Before the execution of an aggregation task, the master issues requests to the controller to launch an aggregation task. The controller must be aware of three kinds of information: 1. The worker numbers of the current aggregation task; 2. The physical topology of the network. Based on these information, the controller constructs an aggregation tree and disseminates this information across the switches. When all the switches receive the information and finishes its configuration, it will send back an acknowledgement to the controller. After ensuring that all the aggregation nodes are rightly configured, the controller replies to the master to start data transmission. Note that the controller is a logical concept, which can be implemented on a dedicated sever, or a middlebox plugged into a switch.

*Server.* The server runs a shim layer which is aimed to exchange information between the workers and the controller. It provides a higher level of abstraction(e.g. GET/PUT interfaces) instead of network interfaces, the worker processes can utilize this abstraction to directly launch aggregation task without consideration of how to communicate with the controller.

## 4. System Design

### 4.1 Network Protocol

*Packet Format.* There are several different kinds of packets transferring through the network. We define and list these packets in Table 1. Each packet contains a traditional L2/L3 routing information (not presented in the table) and a specific packet type. The Launch packet is used to launch an aggregation task, the information of the task is exchanged between the controller and the master server (which is responsible for partitioning data between different workers). The Configure packet is used by the controller to configure the switch for a new aggregation task. A switch manages its memory and commit the aggregation task based on this information. The third type of packets is Ack, which is used for confirmation. Type 0 is used between the controller and master, while type 1 is used between the controller and the switch. The last type is the Aggregation packet, which carries the data that needs to be aggregated. The packet is composed of a TreeID, indicating which aggregation tree it belongs to, an indicator EoT to decide whether it is the last packet of one worker, and several Key-Value pairs of variable lengths, each pair is accompanied with a metadata, which describes the key length and value length.

| Packet Type | Format |
|---|---|
| Launch | <Number of Reducers, Number of Mappers, <List of Reducer Addr>, <List of Mapper Addr>> |
| Configure | <Number of Aggregation Trees, <List of TreeID, Number of Children>> |
| Aggregation | <TreeID, EoT, Operation, Number of Pairs, <List of KeyLength, ValueLength, Key, Value>> |
| Ack, Type 0 / 1 | <NULL> |

Table 1. Packet Type

*Routing.* SwitchAgg utilizes static routing for normal communication and Launch/Configure/Ack packets, the controller is responsible for disseminating the routing table for each switch based on the information of the network topology. For aggregation packets, each switch forwards the aggregated results based on the structure of the aggregation tree, which determines the parent of the aggregation nodes. The construction of aggregation tree is out of the scope of this paper.

### 4.2 Switch Data Plane Design

Figure 4. shows a detailed architecture of the switch. When a packet arrives at the switch, the Header Extraction module first examines the packet header to decide how to handle this packet. If the packet is a normal communication packet, it is passed to the forwarding module for routing, if it is an aggregation packet, it is delivered to Payload Analyzer and enters into the aggregation pipeline. The configure packet will be handled by the configure module.

#### 4.2.1 Header Extraction, Routing and Forwarding

When a packet enters the switch, it first goes through the header extraction module to determine its packet type and how it should be processed further. According to the categories in Table 1, different packets go into different pipelines. If the packet is a normal communication packet, it will be directly forwarded to the routing module for table lookup and decides which output port is should go to. If it is a configure packet, it will be handled by the configure module and the aggregation packet should be passed to Payload Analyzer for further processing. The routing module checks the packet's L2/L3 address and determines which port the packet should go to just as the traditional switches do. The forwarding module forwards different packets based on their packet type and metadata. For communication packet, it forwards it to the proper output queue corresponding to its port number. For aggregation packet, its output port is determined by the configuration tree, hence the forwarding module reads the configuration information and forwards the packet to the proper output queue.

#### 4.2.2 Configuration

The configuration module is responsible for two tasks: 1). Make decisions on how to divide memory for different aggregation trees. 2). Preserve the child number and forwarding port of each tree. For the first task, since we have no detailed information about each tree, we roughly and evenly divide memory among different trees. For example, when a switch has a memory capacity of M, and is owned by two aggregation trees, each tree will occupy half of the memory. For simplicity, in later discussion, we only consider a single tree which owns a part of the total memory, and its base physical memory address can be calculated easily. For the second task, the configuration should maintain those information for processing engine to determine when to flush data to next hop.

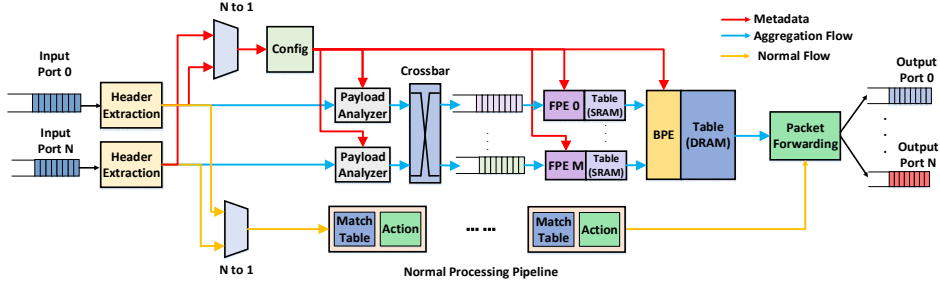

Figure 4. Switch Architecture

## 4.2.3 Payload Analyzer

As we have discussed in section 2, the RMT switch needs to handle Key-Value pairs in a pre-defined format and fixed length, which has several drawbacks and is not flexible enough to deal with aggregation tasks, hence we decide to encapsulate the Key-Value pairs in the payload to achieve both flexibility and efficiency. As shown in Table 1, a packet contains several different-length Key-Value pairs, accompanied with metadata that contains key length and value length. The maximum number of Key-Value pair that a packet can carry is restrained by the maximum packet length and Key-Value length. We have noted a key difference between the Key-Value pair in aggregation task and in normal Key-Value store is that the value in the former task is usually numeric instead of arbitrary string. Hence we consider the value to be a fixed 32-bit integer to simplify our processing.

The greatest challenge we have encountered is how to deal with variable-length keys. As is known before, to handle variable-length keys in Key-Value store is particularly very difficult and inefficient [24], hence designing a processing engine that deals with variable-length Key-Value pairs can hardly meet our performance goal which needs to handle packets at line rate. We take an eclectic approach to deal with this problem. We suppose the length of different key lengths are within the range of [M, N] in the units of Bytes. We define several groups $G_x, G_{x+1}, \ldots, G_{x+k}$, where $x * B < M <= (x + 1) * B$, $(x + k - 1) * B < N <= (x + k) * B$ and B is the base that used to divide the key length range. A key length of L which satisfies that $2x+m < L <= 2x+m+1$ will be divided into group $G_{x+m+1}$. Figure 5(a) shows this division.

After each Key-Value pair has been grouped, they will be transferred through a crossbar, which forwards them to a dedicated processing engine. Each processing engine is responsible for a particular group. Figure. 5(b) presents a division of three groups.

## 4.2.4 Processing Engine

A processing engine is responsible for aggregating the different Key-Value pairs of the same key. Typically, it first searches the key in the memory, then aggregates the value or forwards it based on the searching results. In previous works[14][20], the processing engine can be a match-action unit with a lookup table in the RMT architecture, or a dedicated processor in the NPU-based solution. As we have discussed in section 2, the former approach aggregates data at line rate, but yields a relatively low data reduction ratio while the latter one has a larger memory capacity (e.g. several gigabytes of DRAM) which leads to a higher reduction ratio but gives non-deterministic throughput since it needs to search variable keys and access the relatively slow DRAM. The challenge is how to achieve a high reduction ratio without degrading the throughput.

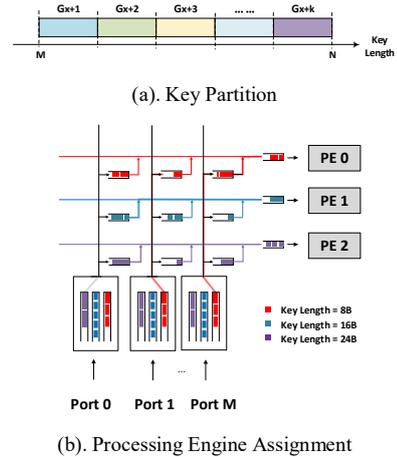

(a). Key Partition

(b). Processing Engine Assignment

Figure 5. Key Range Partition and PE assignment

We take a novel approach to deal with this problem which we call multi-level aggregation hierarchy. Figure 6 shows a logical view of this architecture. The hierarchy is composed of two kinds of processing engines: Front-end Processing Engine and Back-end Processing Engine, we call them FPE and BPE for short. Both FPE and BPE contain a hash function unit, an aggregation unit, and a memory management module accompanied with several amounts of memory (SRAM or DRAM).

We instantiate several FPEs for variable length Key-Value pairs and only one BPE is used for digesting different Key-Value pairs from FPEs. As we have mentioned earlier, each FPE is responsible for a particular range of Key-Value pairs. When a FPE receives a Key-Value pair from the crossbar, it first calculates the hash of the key and looks up the key in its memory, if the key is found, the value is read back and aggregated by the aggregation module. If the key is not found, the Key-Value pair will be stored into the hash table. In the case when hash collision happens, the previously stored key will be evicted and forwarded to the BPE for further processing. A scheduler is sitting between the FPEs and BPE to decide which FPE can forward its result to BPE. Figure 7 presents this process.

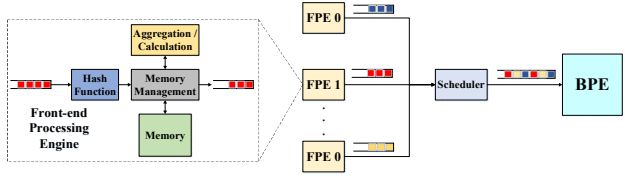

Figure 6. Multi-level Aggregation Hierarchy

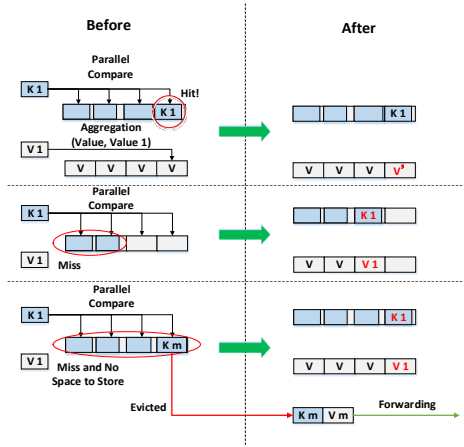

Figure 7. Data Aggregation and Forwarding

This design choice brings us three benefits. First of all, each FPE is only responsible for processing a small range of Key-Value pairs, with a carefully designed hash table, the search and aggrgation can be done in two clock cycles without any pipeline stall. Second, a back-end memory gives us a larger memory capacity, leading to a higher reduction ratio. Third, unlike other implementations such as NPU-based approach, where a cache miss will lead to access to memory which incurs significant delay and degrades throughput, in the multi-level aggregation hierarchy, when a key is missed in the FPE, it will be forwarded to the BPE, where the key processing can be paralleled, this design hides the latency of accessing the relatively slow back-end storage, which facilitates us to digest the network traffic at line rate. We detail the core parts of the Processing Engine.

*Hash Function.* Hash Function module inputs keys and outputs an index, which is used to locate the key. For a given hash function, it can accepts different length inputs and gives a fixed length output. We use the same hash function for different PEs.

*Memory Management and Hash Table.* Memory Management module organizes the memory as a hash table, which is the critical part to store the Key-Value pair. For a contiguous memory space, the memory management module divides them into several hash buckets, and each bucket contains several hash slots. A bucket can be indexed by the hash of the key. To decide whether the key has been stored, all the slots in the same bucket need to be compared to the key. A hash table that belongs to a particular FPE contains hash slots of the same length, where a key has a length less than the hash slot will be padded with zero.

For BPE, it will deal with different length key value pairs, the situation becomes a little different. We divide the memory space into different parts, and each part is corresponding to a particular Group, the memory space of the Group will be arranged in the same manner as the FPE. Figure 8 presents the layout of the hash table in FPE and BPE.

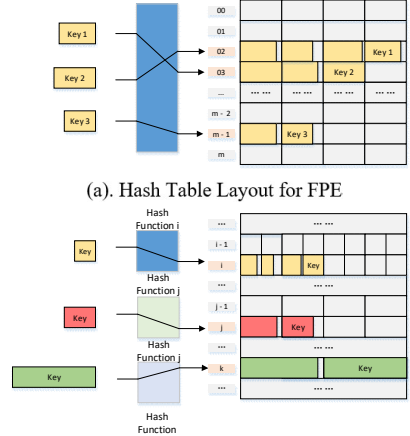

(a). Hash Table Layout for FPE

(b). Hash Table Layout for BPE

Figure 8 Hash Table Layout

**Aggregation Unit.** The aggregation unit accepts the parameters passed from the memory management and returns the aggregated results back to the memory management for further processing. The parameters can be formed as <Operation, Value1, Value2>. We suppose to support different operations including SUM, MAX, MIN, which is frequently used in the aggregation tasks.

## 5. Implementation

We have implemented a prototype of SwtchAgg, including all switch data plane features we described in section 4, a controller for switch configuration and a server agent that provides a shim layer which offers an interface for the workers to launch aggregation tasks.

The switch data plane is developed in Verilog HDL, and compiled into a NetFPGA-SUME development board, which has a Xilinx Vertex-690T chip, accompanied with four 10Gbps SFP+ interface, and 8GB of DRAM with two DDR3 channels. We instantiate 4 payload analyzers, each of them is associated with a dedicated port. To meet the demand of 10Gbps data rate, we configure the interface between different modules to be 128-bit, and runs at a clock frequency of 200Mhz. Different Key-Value pairs will be sent to different front-end processing engines based on their length. We choose to divide the keys into 8 Groups, with an inferior limit of 8B and an upper limit of 64B. Based on this choice, we configure eight FPEs, each of them is responsible for a particular group.

The memory management module accepts notification from the configuration module and decide how to address the memory. For FPE, we chooses on-chip SRAM for key-value storage, for it can be read and written in one clock cycle. For BPE, it is associated with the slower DRAM, with a latency of about 25 clock cycles. Hence we design a memory controller, which buffers the read/write commands and returns the operation results to pipeline the processing. The memory management should be aware of how much tasks it is dealing with in parallel. As we have discussed in previous section, several aggregation trees will share the memory capacity, hence the memory management should be able to address different memory region. The memory management maintains a base pointer for each task, when accessing a particular Key-Value pair, it first decides which memory region it belongs to and then addresses the item according to the base pointer and the index of the key. For BPE, it should also maintain the pointer of different regions of different lengths, the address could be decided as [region base + key range base + key index].

We have also implemented a simple MapRedue-like system, which works in a partition/aggregation pattern. The system is composed of a master and several workers. When launching a new task, the master distributes works among the workers, and select the mapper and reducer nodes. It then communicates with the controller via a shim layer to configure the aggregation tasks. The shim layer and communication library is built on a user-level network stack.

## 6. Evaluation

In this section, we provide evaluation results of SwitchAgg. The results demonstrate that SwitchAgg aggregates data at line rate, and provides significant performance improvements on job completion time for distributed frameworks.

### 6.1 Methodology

*Testbed.* Our testbed consists of 5 server machines. Each server is equipped with two 12-core CPUs (Intel Xeon E5-2658A) and 128GB memory. Three machines are used as mappers to generate Key-Value pairs that need to be aggregated. One machine is plugged with a NetFPGA-SUME card through which has four 10Gbps SFP+ interface, we implement our switch processing pipeline on this card. Another machine is used as reducer to generate the final result, the controller and master are also implemented on this server. Mappers and Reducer are directly connected to different physical ports of the switch,

*Workloads.* Each mapper generates key-value pairs of different size. The range of key length is 16B ~ 64B. Two parameters are used for different configuration. Workload size determines the total amounts of data we will geneate and Memory capacity determines the amount of BRAM memory we use in Front-end Processing Engine. The workload size ranges from 2GB to 16GB and the memory capacity of FPE BRAM ranges from 4MB to 32MB. We fix the key variety to 1GB, which represents a typical case in MapReduce job. We use both uniform and skewed workloads. The skewed workloads follow Zipf distribution with skewness parameter of 0.99. Each mapper generates workloads based on these two parameters, in each test, the mappers share the same parameters. To avoid disk I/O overheads, all the data is stored in memory.

### 6.2 SwitchAgg Micro-benchmark

We first show switch micro-benchmark results to illustrate how SwitchAgg can significantly reduce network traffic in aggregation tasks. And then we show that SwitchAgg processes data at line rate.

*Reduction ratio vs. Workload size/Memory Capacity.* Three mappers generate workload based different parameter configurations, and then send these key-value pairs to the reducer. We add counters in the switch ports to measure the amount of input data and the output data, and then we calculate the traffic reduction ratio. Figure 9 illustrates the results, S-{4, 8, 16, 32}MB represents different sizes of BRAM, and M-32MB means multi-level aggregation.

Before we turned on Multi-level Aggregation, the reduction ratio is below 5% in the uniform case, since the BRAM cannot deal with the large amounts of key variety. Even if we increase the memory capacity to 32M, the reduction ratio is still below 10%. In the case of Zipf distribution, the reduction ratio is relatvely higher because all of the hot keys can be aggregated in the Front-end engine. However, as we have observed, after increasing the total amounts of workload, the reduction ratio is lower because some of the hot keys cannot be aggregated and they contribute a lot to the output traffic. After we enable the Multi-level aggregation mechanism, the reduction ratio is rising. Since the two-level memory hierarchy can accommodate almost all the keys. In the highly skewed case, the reduction ratio can even reach 99% or higher.

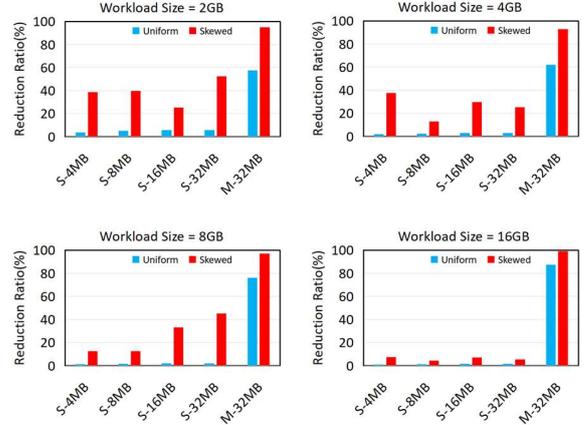

Figure 9. Reduction ratio

*Aggregate at line rate.* One of the most important measurement of our prototype is whether it can relay packets at line rate. However, since the packet is digested in the switch, we cannot simply measure the output rate and compare it to the input rate. Hence we take a different approach to see whether a packet is digested in the processing engine at line rate. Each processing engine reads data from a FIFO, we set two counters to measure how many times the FIFO is written into and how many times the FIFO is full, which means the processing engine is not able to process at line rate. Table 2 gives the results.

| Workload size | Written Times | FIFO-Full times | Full-time ratio |
|---|---|---|---|
| 2GB | $6.22 * 10^7$ | $2.74 * 10^4$ | 0.044% |
| 4GB | $11.93 * 10^7$ | $3.81 * 10^4$ | 0.032% |
| 8GB | $22.43 * 10^7$ | $6.25 * 10^4$ | 0.028% |
| 16GB | $48.29 * 10^7$ | $20.58 * 10^4$ | 0.042% |

Table 2. FIFO-Full Time Ratio

The figure shows that less than 1% of the packets is waiting for the former one to be processed. The waiting time is due to hash collision and forwarding to the back-end processing engine. This result shows that the processing engine is able to digest packet at line rate.

***Transmitting Delay.*** The high data reduction ratio is not for free. The process of aggregation incurs additional delay, we divide the extra latency of the processing pipeline. Table 3 shows this division.

| Stage | Delay(Cycles) |
|---|---|
| Header Analyzer | 3 |
| Crossbar | 2 |
| FPE-Hash | 10 |
| FPE-Aggregate | 18 |
| FPE-Forward | 5 |
| BPE-Aggregate | 33 |
| BPE-Flush | $3.125 * 10^7$ |

Table 3. Processing Delay

We observe that the main contribution of latency is the processing engine, which needs to flush data from the memory to the next stage, since we run at a clock rate of 200MHz, the flush stage takes nearly 78ms. However, as we will show in the following section, this delay is negligible when compared to the overall performance of job completion time.

### 6.3 System Performance

We evaluate how SwitchAgg can improve the overall system performance. We run a Word-Count instance on the mappers and reducers, which is a typical example of MapReduce. We generate 4 workloads of different size, ranging from 2GB to 16GB. The key variety is fixed to 1GB, and we use highly skewed key distribution since the word distribution usually follows a Zipf distribution. We both enable and disable Multi-level processing. We then see how long it will take to complete a job, and how much impact it will have in CPU utilization.

***Job Completion time.*** Figure 10 shows the job completion time with and without SwitchAgg assistance. We find that, the more workload we have, the more time SwitchAgg can save. Which reflects that a higher data reduction ratio helps to reduce the job execution time a lot. However, in some cases, we also find that the result of with- and without SwitchAgg is similar, this is because the overhead of SwitchAgg offsets its benefits. When we get a maximum workload of 16GB, the completion time can be reduced to more than 50%.

***CPU utilization.*** SwitchAgg not only brings us the benefits of smaller job completion time, it can also help us to reduce CPU consumption, which is a vital resource in cloud environment. Figure 11 illustrates the average CPU utilization during the job execution. An obvious conclusion we can draw from the results is that the higher the data reduction ratio is, the lower the CPU utilization is. The saved CPU time can be used to execute other job and improve the overall performance.

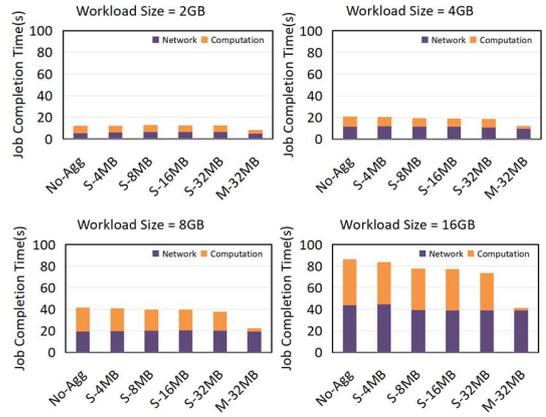

Figure 10. Job Completion Time

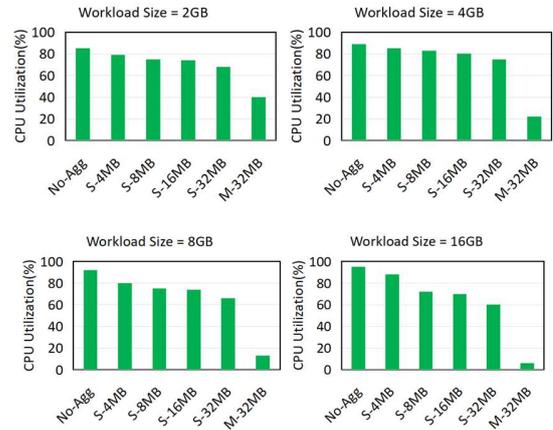

Figure 11. CPU Utilization

## 7. Discussion and Future Work

SwitchAgg is an in-switch aggregation system, which we believe will be an important building block for a general in-network processing system. The key observation in this work is that by carefully design the processing and memory hierarchy of the switch, we can achieve fast and high-performance datacenter workload processing. There still exists many questions left to be discussed, to this end, we plan to explore the following topics in the future.

***Performance Modeling.*** Traditional performance model for distributed frameworks, such as LogP [25] model, usually takes the network as a black box, as it is only responsible for transferring the data. However, as the emerging trends of programmable devices, the network is much more powerful, which has the ability to participate in data processing, and will in turn affect the performance modeling and analysis. We plan to design a new model to take network processing into consideration to complement the existing methods.

*Network Routing Scheme.* Traditional network routing is based on a premise that the input and output traffic is roughly the same, however, in-network computation will significantly change this case. In the aggregation scenario, a switch which digests multiple flows may output little traffic, which may affect the routing of other non-aggregated flows.

*Memory Utilization.* As we have discussed before, two aggregation trees which share a same switch will evenly divide the memory. However, this may not be an optimistic solution, if the one aggregation tree has much more data needs to be aggregated. We suppose the application can provide more information to guide us utilize the switch memory much more efficiently. Furthermore, next-generation switch may support multiple in-network aggregation tasks, how to efficiently manage the memory is also a great challenge.

## 8. Related work

Camdoop [13] is a system which supports on-path aggregation for MapReduce-based applications. It requires a custom topology where servers are directly connected to each other. Thus, it is incompatible with a common data center infrastructure.

NetAgg [11] utilizes middle-boxes to commit on-path aggregation. The middle-boxes are connected to switches through high-bandwidth links. When a switch receives a packet that needs to be aggregated, it first forwards it to the middle-box and then receives the results and forwards them. Similar to Camdoop, this method also requires changes in the network architecture, besides, this software-based approach cannot provide high-performance data aggregation, which can become a performance bottleneck.

SHArP [15] is designed to accelerate traditional scientific applications. It is mainly responsible for offloading MPI collective operation processing to the network. SHArP offers high throughput and low latency since it is implemented in the Switch ASIC, however, this implementation also constrains it be widely used in datacenter environment.

DAIET [14] utilizes the programmable switch to implement in-network data aggregation. They provide a high data reduction rate without need to modify the network infrastructure. However, as we have discussed before, DAIET cannot maintain a high data reduction rate while there is a large volume of data needs to be aggregated.

Unlike NetAgg and DAIET, SwitchAgg does not modify the network architecture and provides both high processing ability with a considerable data reduction rate.

## 9. Conclusion

In this paper, we first give a detailed analysis of the drawbacks of current in-network aggregation implementations. Based on this analysis, we propose our design of SwitchAgg, prototype it and evaluate its performance. The experimental results show that our system can process in-network aggregation tasks at line-rate while maintaining a high data reduction rate. This result demonstrates a very promise future of in-network computation. Besides, we also give several advises to direct future designs, we believe our work to be a concrete step towards general-purpose in-network computation.